\documentclass[iop]{emulateapj}

\slugcomment{Draft version March 12, 2018}

\shorttitle{Mass Modeling of MACS1149}
\shortauthors{Finney et al.}

\begin{document}


\title{Mass Modeling of Frontier Fields Cluster MACS J1149.5+2223 \\
    Using Strong and Weak Lensing}


\author{Emily Quinn Finney\altaffilmark{1}} 
\author{Maru\v{s}a Brada\v{c}\altaffilmark{1}}
\author{Kuang-Han Huang\altaffilmark{1}}
\author{Austin Hoag\altaffilmark{1}}
\author{Takahiro Morishita\altaffilmark{2}\altaffilmark{3}\altaffilmark{4}}
\author{Tim Schrabback\altaffilmark{5}}
\author{Tommaso Treu\altaffilmark{2}}
\author{Kasper Borello Schmidt\altaffilmark{6}}
\author{Brian C. Lemaux\altaffilmark{1}}
\author{Xin Wang\altaffilmark{2}}
\author{Charlotte Mason\altaffilmark{2}}


\email{Email: eqfinney@ucdavis.edu}
\altaffiltext{1}{Department of Physics, University of California,
    Davis, CA 95616, USA}
\altaffiltext{2}{Department of Physics and Astronomy, 
  University of California (UCLA), Los Angeles, CA, 90095, USA}
\altaffiltext{3}{Astronomical Institute, Tohoku University,
  Aramaki, Aoba, Sendai 980-8578, Japan}
\altaffiltext{4}{Institute for International Advanced Research and Education,
  Tohoku University, Aramaki, Aoba, Sendai 980-8578, Japan}
\altaffiltext{5}{Argelander-Institut f\"{u}r Astronomie, 
    Auf dem H\"{u}gel 71, D-53121 Bonn, Germany}
\altaffiltext{6}{Leibniz-Institut f\"{u}r Astrophysik Potsdam (AIP),
  An der Sternwarte 16, 14482 Potsdam, Germany}


\begin{abstract}
We present a gravitational lensing model of MACS J1149.5+2223 using
ultra-deep Hubble Frontier Fields imaging data and spectroscopic redshifts 
from HST grism and VLT/MUSE spectroscopic data. We create total mass 
maps using 38 multiple images (13 sources) and 608 weak lensing galaxies, as 
well as 100 multiple images of 31 star-forming regions in the galaxy that hosts
Supernova Refsdal. We find good agreement with a range of recent models within 
the HST field of view. We present a map of the ratio of projected stellar mass 
to total mass ($f_{\star}$), and find that the stellar mass fraction for this 
cluster peaks on the primary BCG. Averaging within a radius of 0.3 Mpc, we 
obtain a value of $\langle f_{\star} \rangle = 0.012^{+0.004}_{-0.003}$, 
consistent with other recent results for this ratio in cluster environments, 
though with a large global error (up to $\delta f_{\star} = 0.005$) 
primarily due to the choice of an IMF. We compare values of $f_{\star}$ and 
measures of star formation efficiency for this cluster to other Hubble Frontier 
Fields clusters studied in the literature, finding that MACS1149 has a higher 
stellar mass fraction than these other clusters, but a star formation 
efficiency typical of massive clusters.
\end{abstract}


\section{Introduction}
Galaxy cluster mass models are among the most useful applications of 
the theory of gravitational lensing to early-universe cosmology. Especially
at redshifts near $z\sim0.5$, massive galaxy clusters have wide areas of large 
magnification power, and thus have significant implications for observing 
faint, high-redshift background objects \citep[][and references 
therein]{bradac14, mohammed16}. Magnification modeling is needed to constrain 
the faint end of the $z=6-8$ rest-frame UV \citep[][and references 
therein]{ishigaki17, livermore17, bouwens16, robertson15} luminosity functions. 
At lower redshifts ($z=0.7-2.3$), galaxy evolution studies have used 
gravitational lensing to spatially resolve galaxy kinematic properties and 
chemical abundances \citep[][and references therein]{mason16, wang16, vulcani16, 
leeth16, jones15, christensen12, richard11, jones10, stark08}. 

Recent analyses have determined that, in extreme cases, differences in 
magnification models can be responsible for systematic uncertainties up to 
several orders of magnitude for lensed sources as faint as $-12$ mag 
\citep{bouwens16}. Uncertainties in magnification have 
significant implications on estimates of the faint end of the luminosity 
function \citep{bouwens16}, underlining the necessity of producing 
magnification maps using a variety of independent techniques to quantify the 
effects of model systematics \citep{priewe17}. The majority of cluster 
magnification maps use parametric modeling techniques (i.e, techniques that 
assume an underlying relationship between dark and luminous matter), which 
typically share the same class of systematic biases. Thus, it is critical to 
produce additional magnification maps using free-form techniques 
\citep{mohammed16}.

In addition to their value in producing magnification maps, galaxy 
cluster mass models yield insight into a number of dark matter properties, 
including the key assumption that light traces mass (LTM). The role 
of dark matter in cluster physics is still poorly understood, and is thus a 
large source of systematic error in simulations of cluster physics 
\citep{dubois13, laporte15, martizzi14, mohammed16}. Current theoretical models 
underpinning cluster simulation studies, such as the cluster subhalo mass 
function \citep[][and references therein]{natarajan17, mohammed16, atek15}, 
depend on the validity of the assumption of LTM. Without testing this claim, 
it is also impossible to investigate alternative theories that challenge 
$\Lambda$CDM, such as dark matter self-interaction \citep{markevitch04, 
clowe06}. Thus, mass models that avoid LTM assumptions are critical for 
improving the theoretical models underlying cluster simulations.

Additionally, the distribution of cluster baryonic matter can reveal 
whether cluster baryons are representative of universal baryonic material 
\citep[e.g.,][and references therein]{gonzalez13, nagai07, simionescu11}, and 
may provide explanation for how efficiently stellar material is formed in 
clusters by comparing the stellar material in clusters to the cluster gas mass 
\citep[][among others]{gonzalez13, behroozi13, kravtsov14} and total baryon 
fraction \citep{gonzalez13, lin03, budzynski14}. Investigations 
of star formation efficiency have implications for simulations of galaxy 
evolution \citep[][and references therein]{gonzalez13} in cluster contexts, 
particularly of the role of radiative cooling and/or stellar feedback 
in heating the ICM and altering its metallicity \citep{dolag09, ettori06, 
kravtsov05, delucia04, valdarnini03}.

One way to address these questions of cluster composition is to 
compare the distribution of the ratio of stellar mass to total mass within 
galaxy clusters. \citep[][and references therein]{bahcall95, bahcall14}. 
Several authors have calculated this ratio for many clusters 
and over a range of distance scales \citep[e.g.][]{bahcall14, gonzalez13, 
wang16, hoag16, jauzac12, jauz16}. \citet{bahcall14} have found that on scales 
larger than a few kpc, the ratio of stellar mass to total mass is roughly 
constant, suggesting that stellar mass traces total mass when averaged over 
large apertures. Galaxy cluster mass models are critical to producing these
results \citep{mohammed16, wang16, hoag16}. 

The galaxy cluster MACS1149.5+2223 ($z=0.544$, MACS1149 hereafter) has been
extensively studied due to its complex merging properties and potential as a 
lensing cluster. It was first identified as part of the MAssive Cluster Survey 
\citep[MACS:][]{ebeling01}, and has been subsequently observed in the optical
band by the Cluster Lensing And Supernova survey with Hubble
\citep[CLASH:][]{postman12} and the Hubble Frontier Fields initiative
\citep[HFF:][]{PI Lotz, 
coe15},\footnote{\url{http://www.stsci.edu/hst/campaigns/frontier-fields/FF-Data}}
as well as by observations in other wavelengths. As a result, the mass
distribution of MACS1149 has been modelled with a variety of lens-modeling
techniques \citep{smith09, zitrin09, zitrin11, richard14, coe15, johns14, 
rau14, mohammed16}, which give broadly similar cluster mass distributions.

The discovery of Supernova Refsdal \citep[$z_s=1.49$,][]{kelly14, kelly16}
in the center of MACS1149 is of particular interest to the modeling community,
as it demonstrated the possibility of using clusters as lenses to magnify
multiply-imaged background supernovae. While lensed supernovae had previously 
been predicted \citep{refsdal64} and observed \citep{rodney15a}, the time 
delay between SN Refsdal's multiple images provided a valuable opportunity to 
test the accuracy of mass models \citep{treu16, snj16, jauz16, diego15, 
grillo15a} in the center of the cluster, and the predicted magnifications of 
these images can be compared to observed flux ratios to provide a tight 
constraint on the model near the BCG. 

In this paper, we present a strong and weak gravitational lensing model of
MACS1149 using HFF data by modeling the cluster with our Strong and Weak
Lensing United (SWUnited) method \citep{bradac09, bradac05}. Unlike the 
majority of lens modeling methods previously used to study MACS1149, 
SWUnited does not explicitly fit the parameters of an underlying dark matter 
distribution that traces stellar light. Indeed, this is the only mass 
map of MACS1149 to date that is produced with both HFF data and a 
non-parametric (gridded) modeling approach. We compare our results to recent 
models of strongly lensed systems in MACS1149 presented by 
\citet{snj16, diego15, grillo15a, jauz16, oguri15, kawamata15, treu16}. 

We also present a map of the stellar mass to total mass ratio, the 
first of its kind to be produced for MACS1149. Since MACS1149 is a massive and 
dynamically complex cluster, it is interesting to consider how well its stellar 
material traces its dark matter. We discuss the distribution of this value 
across the cluster, and its implications for estimates of star formation 
efficiency.

The paper proceeds as follows: in section 2, we describe in detail the data
used for spectroscopic and photometric redshift identification, as well as for
weak lensing. Section 3 discusses the process of modeling MACS1149 using the
SWUnited method and presents modeling results; a stellar mass to total mass
ratio map is presented in section 4. Finally, section 5 concludes these
results. 

Throughout the paper, we adopt a cosmology with $\Omega_{m} = 0.3$, 
$\Omega_{\Lambda} = 0.7$, and $h=0.7$. We give all magnitudes in the 
AB system \citep{oke74}.

\section{Data and Image Identification}
To constrain our mass model, we use imaging data to identify strongly-lensed
multiple images and weakly-lensed background galaxies, and to estimate their 
redshifts. We use spectroscopic data to more accurately estimate multiple 
image redshifts \citep{treu15}, since lens models tend to substantially 
improve with more spectroscopic data \citep{rodney15a, rodney15b}. 

\subsection{HST and VLT Spectroscopic Data}
Original spectroscopy for MACS1149 was obtained from the Grism 
Lens-Amplified Survey from Space \citep[GLASS:][]{schmidt14, 
treu15}\footnote{\url{http://glass.astro.ucla.edu}}$^,$\footnote{\url{https://archive.stsci.edu/prepds/glass/}},
which was initiated to study highly-magnified high-redshift objects.
The GLASS collaboration obtained 14 HST orbits per cluster, with the 
WFC3-IR G102 (10 orbits) and G141 (4 orbits) grisms (field of view
$123\times136\farcs$). Each observation was taken at two position angles 
$90^{\circ} \pm 10^{\circ}$ apart to improve deblending and spectrum extraction.
The extraction apertures were $0\farcs6\times100\AA$ (5 spatial by 3 
spectral native pixels). The average spectral resolution was $R\sim130$ for the 
G141 grism, and $R\sim210$ for the G102 grism. Overall, the spectra covered a 
continuous wavelength range of $0.81-1.69\mu$m. Direct image exposures were taken 
in the F105W and F140W bands to ensure image alignment and calibration 
\citep{treu15}. Deep HST G141 data were taken as part of the Supernova Refsdal 
follow-up campaign \citep[HST-GO-14041; PI Kelly][]{brammer16}, adding 
another 30 orbits of G141 spectra.

In addition to the HST spectroscopy, spectroscopic data were taken with 
VLT/Multi Unit Spectroscopic Explorer \citep[MUSE,][]{bacon10}; these 
observations were designed to measure multiple image redshifts and to observe 
SN Refsdal \citep{karman16, kelly16, grillo15a, rodney15a}. Observations were 
taken in the wavelength range $480-930$ nm, with a total integration time of 
4.8 hours. The field of view was $1\times1\farcm$, with an average spectral
resolution of $R\sim3000$. Most spectroscopic redshifts of multiply-imaged 
objects in this work were obtained from the VLT/MUSE data due to its sensitivity
and wavelength coverage being more favorable than those of GLASS for the sources
in question. Further details of these observations and analyses are discussed 
in \citet{grillo15a, kelly16}. 

\subsection{HST and Spitzer Imaging Data}
As part of the HFF initiative (HST-GO 13504, PI J. Lotz), MACS1149 was 
observed in seven filters: three ACS/optical filters (F435W, F606W, F814W) 
and four WFC3/IR filters (F105W, F125W, F140W, F160W). These data were 
taken between November 2014 and June 2015 (141 orbits), and were supplemented 
by additional WFC3/NIR data from the SN Refsdal follow-up campaign (20 orbits,
PI Rodney HST-GO 13790, PI Kelly HST-GO 14041). 
We use both epochs of the public release data products from the HFF team, 
yielding a $5\sigma$ point source sensitivity of $\sim 29$mag per band. 
Further supplementary data were obtained from the CLASH program \citep[20 
orbits,][]{postman12}, and were primarily used to construct our photometric 
catalog. 

We use deep \emph{Spitzer}/Infrared Array Camera (IRAC) images in combination 
with the HST images to improve stellar mass and photometric redshift 
measurements of cluster members and background galaxies. The final 
\emph{Spitzer}/IRAC imaging mosaic, made available by P. Capak, includes data 
taken by the IRAC Lensing Survey (PI: Egami), \emph{Spitzer} UltRa Faint 
SUrvey Program \citep[SURFSUP:][]{bradac14, ryan14, 
h16b},\footnote{\url{http://bradac.physics.ucdavis.edu/\#surfsup}} and the 
\emph{Spitzer} Frontier Fields program (Capak et al., in 
prep).\footnote{\url{http://irsa.ipac.caltech.edu/data/SPITZER/Frontier/\\
images/MACS1149/}} The mosaic is comprised of ten dithered 
$5\farcm2\times5\farcm2$ images, with a total field of view of 
$20\farcm6\times29\farcm6$ ($7.9\times11.3$ Mpc). The combined exposure time 
reaches $\sim$50 hours in each of channel 1 ($3.6\mu$m) and channel 2 
($4.5\mu$m). The images reach a $5\sigma$ limiting magnitude of 
$\sim$26.5 mag (channel 1) and $\sim$26.0 (channel 2) for a point source 
outside the cluster core, sensitive enough to probe the bulk of cluster 
member stellar mass. 

\subsection{Photometry}
The majority of the background sources do not have spectroscopic redshifts,
so their redshifts are estimated using photometric measurements from HST
images. We follow the procedure by \citet{h16b} to measure broadband flux
densities. In short, we first convolve all images blueward of F160W 
with PSF smoothing kernels to match the PSF width in F160W. We do this 
by using the \texttt{psfmatch} task in \texttt{IRAF} to compute the PSF kernels
on individual point sources in the cluster field. We measured 
the FWHM of the PSF in F160W and found that it is around $0\farcs18-0\farcs2$. We 
also compared the curves of growth using point sources across different filters 
and found that after convolving with PSF-matching kernels, the curves of growth 
generally agree with that in F160W to within $15\%$. We then measure 
broadband flux densities using SExtractor \citep{bertin96} in dual-image mode, 
with the F160W image as the detection image. The total flux density of each 
source is measured within an elliptical Kron aperture with $k=2.5$ 
\citep[][called \texttt{MAG\_AUTO} in SExtractor]{kron80}, and the colors are 
calculated using the flux densities measured within isophotal apertures 
(called \texttt{MAG\_ISO} in SExtractor).

Photometric redshift of each source is estimated using the colors in HST
and \textit{Spitzer} filters. We use the photometric redshift code EAZY
\citep{brammer08} which employs a template-fitting approach using a set of
templates that are optimized for redshift estimation. The code produces
probability distribution functions for the redshifts of each source; we take
the peak of this distribution as our photometric redshift estimate and use the
distribution to derive our confidence intervals. In our redshift fitting
process, we do not employ a redshift prior, and a comparison with known
spectroscopic redshifts shows that the photometric redshifts are consistent
with spectroscopic redshifts within error bars. 


\subsection{Weak Lensing Catalog}
We generate a weak lensing catalog based on a 26.9ks deep stack of ACS F606W 
imaging using the pipeline described in \citet{schrabback18a}. This 
pipeline utilizes the \citet{erben01} implementation of the KSB+ algorithm
\citep{kaiser95b, luppino97, hoekstra98} for galaxy shape measurements as 
detailed in \citet{schrabback07}. It additionally employs the pixel-level 
correction for charge-transfer inefficiency from \citet{massey14}, as well 
as a correction for noise-related shape measurement biases and a principal 
component model for the temporally and spatially varying ACS point-spread 
function \citep{schrabback10}. \citet{schrabback18b} and 
Hern{\'a}ndez-Mart{\'{\i}}n et al. (in prep) have extended earlier 
simulation-based tests of the employed shape measurement pipeline to the 
non-weak shear regime of clusters (for \mbox{$|g|\le 0.4$}), confirming that 
residual multiplicative shear estimation biases are small 
(\mbox{$|m|\lesssim 5\%$}). For the source selection and redshift estimation 
we follow \citet{hoag16}, cross-matching objects in our weak lensing catalog 
with those in our photometric catalog produced from CLASH data.

\subsection{Strong Lensing Catalog}
We use strong lensing images to constrain our mass model. In this work, 
we use the grading system of the \citet{treu16} modeling collaboration
\citep{wang16, hoag16, treu16}, identifying multiple images as being part of
the same system using color, morphology, and spectral emission features. Each
potential multiple image system was inspected by five independent teams of the
HFF collaboration and identified as gold (very likely to be images from the
same system) or silver (likely to be images from the same system)
based on image colors and morphologies. Any systems not classified as gold or
silver are not likely to be multiple images, and they are not used in this
model. 

A total of 23 images (8 systems) are spectroscopically confirmed by 
\citet{treu16}, and are listed in Table 1. In system 13, the images have somewhat different 
spectroscopic redshifts despite being consistent morphologically with each 
other. The spectra were obtained from HST grism observations with a 
$68\%$ confidence interval of $\pm 50 \AA$, which corresponds to a 
$\delta_z \approx 0.01$. One image has a spectroscopic redshift of 
$z=1.25\pm0.01$ and the other two have redshifts of $z=1.23\pm0.01$, so for 
this system the average of the individual spectroscopic redshifts was used as 
the system redshift. Because of the strong morphological consistency and correct 
parity of the system, it is considered to be a true system. A second system 
(system 5) had only one of its images spectroscopically confirmed; the other 
images are considered to share the same redshift based on their colors and
morphologies. 

\begin{figure}
  \includegraphics[width=0.49\textwidth]{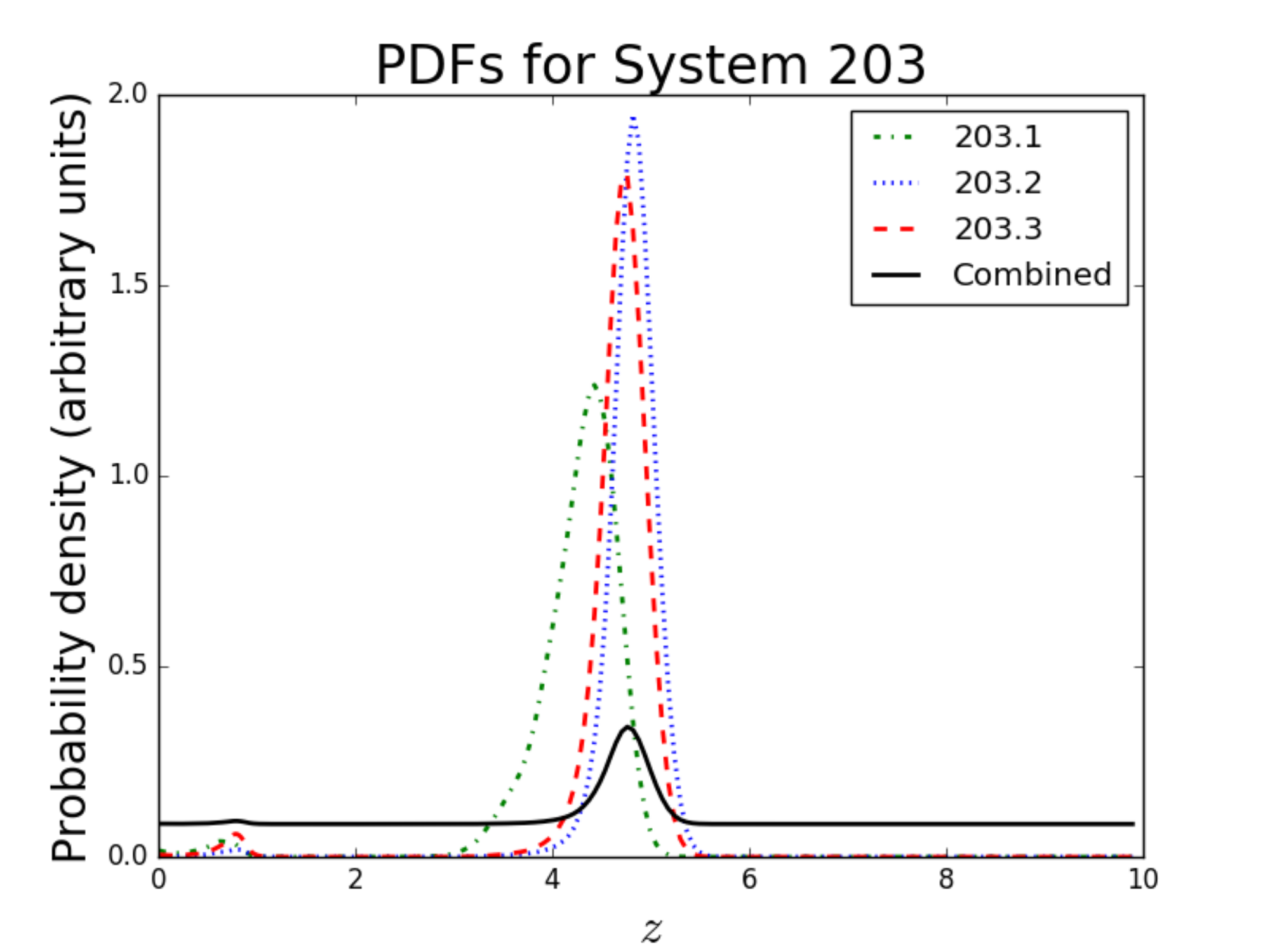}
  \caption{\label{fig:PDFs} Example of a normalized redshift probability
    density function, P($z$) for multiple images in a photometric system. 
    The bold black line shows combined redshift probability density
    functions for multiple images in photometric systems, as obtained by
    our Bayesian combination procedure.}
\end{figure}

We supplement the spectroscopically-confirmed multiple images with 15 
images (5 systems) identified photometrically as strong multiple image 
candidates. An example of a photometric redshift P($z$) for a strongly-lensed 
system is shown in Figure \ref{fig:PDFs}. For multiple images not confirmed by 
spectroscopy, the model presented in this work includes only systems with at 
least two images that have coinciding peaks (i.e., peaks are within each 
other's 68\% confidence intervals) in redshift probability density functions, 
P($z$), and whose most likely photometric redshifts are greater than the 
cluster redshift. Redshift estimates used in the mass model are 
listed in Table 1. Some of the multiple images are blended with 
foreground sources; thus, their redshifts cannot be reliably estimated and we 
remove their incorrect P($z$) from our strong lensing redshift estimates.

In addition to using these multiple images of sources in independent 
redshift planes, we further constrain the model using 100 images of 31 unique 
star-forming regions present in the highly-resolved spiral galaxy that hosts 
SN Refsdal \citep[Table 2,][]{treu16}. Several images of this galaxy appear near 
the cluster center, and the number and quality of these associated multiple 
images provides a powerful constraint to the mass model in these central 
regions. All of the images of the star-forming regions in the SN Refsdal host 
galaxy were included in several models \citep{diego15, kawamata15, sharon15, 
grillo15a}, and most were included by \citet{grillo15a}, Zitrin et al. (in 
prep) as well. 

Of the 138 images (44 systems) used in the model, 123 images (39 
systems) have spectroscopically measured redshift and only 15 images (5 
systems) are identified photometrically, so we are confident in the robustness 
of our strong lensing catalog.

\clearpage
\LongTables 
\begin{deluxetable*}{llllllllllll}
\tablecolumns{12}
\tablewidth{0pc}
\tablecaption{Multiply Imaged Systems \label{tab:arcs}}
\tablehead{        
\colhead{ID} & 
\colhead{$\alpha$} & 
\colhead{$\delta$} & 
\colhead{Spec-$z$} & 
\colhead{Peak} &
\colhead{System} &
\colhead{Magnification} &
\colhead{Category} \\
\colhead{} & 
\colhead{(J2000)} & 
\colhead{(J2000)} & 
\colhead{} & 
\colhead{P($z$)} & 
\colhead{Redshift} &
\colhead{} &
\colhead{}
}
\startdata
1.1&   177.39700& 22.39600& 1.488   &\nodata             &1.488               &$ 12 \pm 1$&Gold \\
1.2&   177.39942& 22.39743& 1.488   &\nodata             &\nodata             &$  7.3 \pm 0.5$&Gold \\
1.3&   177.40342& 22.40243& 1.488   &$1.5^{+0.1}_{-0.3}$ &\nodata             &$  3.90 \pm 0.04$&Gold \\
1.5&   177.39986& 22.39713& \nodata &\nodata             &\nodata             &$ 13 \pm 2$&Silver \\\cline{1-12}
2.1&   177.40242& 22.38975& 1.891   &$2.3^{+0.2}_{-0.1}$ &1.891               &$  4.9_{- 0.2}^{+  0.1}$&Gold \\
2.2&   177.40604& 22.39247& 1.891   &\nodata             &\nodata             &$ 35_{-1}^{+ 23}$&Gold \\
2.3&   177.40658& 22.39288& 1.891   &$2.4\pm 0.1$        &\nodata             &$ 13 \pm 2$&Gold \\\cline{1-12}
3.1&   177.39075& 22.39984& 3.129   &$3.2\pm 0.1$        &3.129               &$ 16 \pm 1$&Gold \\
3.2&   177.39271& 22.40308& 3.129   &\nodata             &\nodata             &$ 13_{- 1}^{+  2}$&Gold \\
3.3&   177.40129& 22.40718& 3.129   &$3.4 \pm 0.1$       &\nodata             &$  6.0 \pm 0.1$&Gold \\\cline{1-12}
4.1&   177.39300& 22.39682& 2.949   &\nodata             &2.949               &$ 12.5_{- 0.4}^{+  0.5}$&Gold \\
4.2&   177.39438& 22.40073& 2.949   &\nodata             &\nodata             &$  8.8 \pm 0.2$&Gold \\
4.3&   177.40417& 22.40612& 2.949   &$2.6^{+0.1}_{-0.2}$ &\nodata             &$  4.3_{- 0.2}^{+  0.1}$&Gold \\\cline{1-12}
5.1&   177.39975& 22.39306& 2.80    &\nodata             &2.80                &$ 26_{- 10}^{+  7}$&Gold \\ 
5.2&   177.40108& 22.39382& \nodata &$2.6 \pm 0.1$       &\nodata             &$  9_{- 1}^{+  6}$&Gold \\
5.3&   177.40792& 22.40355& \nodata &$2.8 \pm 0.1$       &\nodata             &$  5.0_{- 0.4}^{+  1.0}$&Silver \\\cline{1-12}
6.1&   177.39971& 22.39254& \nodata &\nodata             &$2.65_{-1.81}^{+3.75}$&$ 10_{- 4}^{+  6}$&Gold \\ 
6.2&   177.40183& 22.39385& \nodata &$2.6^{+0.1}_{-2.3}$ &\nodata             &$  7.0_{- 0.5}^{+  1.1}$&Gold \\
6.3&   177.40804& 22.40250& \nodata &\nodata             &\nodata             &$  5.5_{- 0.1}^{+  0.2}$&Silver \\\cline{1-12}
7.1&   177.39896& 22.39133& \nodata &$2.6 \pm 0.1$       &$2.56_{-1.69}^{+3.84}$&$  6.3_{- 0.8}^{+ 18.0}$&Gold \\
7.2&   177.40342& 22.39426& \nodata &$2.7^{+0.1}_{-0.2}$ &\nodata             &$  5.2 \pm 0.2$&Gold \\
7.3&   177.40758& 22.40124& \nodata &$2.3^{+0.1}_{-0.2}$ &\nodata             &$  4.4_{- 0.6}^{+  0.2}$&Gold \\\cline{1-12}
13.1&  177.40371& 22.39778& 1.23    &$1.3 \pm 0.1$       &1.24                &$ 12.1_{- 0.9}^{+  0.7}$&Gold \\
13.2&  177.40283& 22.39665& 1.25    &\nodata             &\nodata             &$ 11.5_{- 0.3}^{+  0.4}$&Gold \\
13.3&  177.40004& 22.39385& 1.23    &$0.8^{+0.2}_{-0.1}$ &\nodata             &$  5.1_{- 0.5}^{+  0.3}$&Gold \\\cline{1-12}
14.1&  177.39167& 22.40348& 3.703   &$3.4^{+0.2}_{-0.1}$ &3.703               &$ 17_{- 1}^{+  2}$&Gold \\
14.2&  177.39083& 22.40264& 3.703   &$3.4 \pm 0.1$       &\nodata             &$ 43_{- 5}^{+  7}$&Gold \\\cline{1-12}
110.1& 177.40014& 22.39016& 3.214   &$3.2^{+0.1}_{-0.2}$ &3.214               &$  6.4_{- 1.0}^{+  0.3}$&Gold \\
110.2& 177.40402& 22.39289& 3.214   &\nodata             &                    &$  5.4_{- 0.5}^{+  0.2}$&Gold \\\cline{1-12}
26.1&  177.41035& 22.38874& \nodata &$3.3 \pm 0.1$       &$3.29_{-2.40}^{+3.06}$&$ 38_{-21}^{+ 37}$&Silver \\
26.2&  177.40922& 22.38769& \nodata &$3.1^{+0.1}_{-0.2}$ &\nodata             &$ 74_{-72}^{+172}$&Silver \\
26.3&  177.40623& 22.38536& \nodata &$3.4 \pm 0.1$       &\nodata             &$  5.8_{- 0.3}^{+  0.4}$&Silver \\\cline{1-12}
203.1& 177.40995& 22.38724& \nodata &$4.4^{+0.2}_{-0.5}$ &$4.70_{-3.73}^{+1.37}$&$ 49_{- 2}^{+ 50}$&Silver \\
203.2& 177.40657& 22.38451& \nodata &$4.8 \pm 0.2$       &\nodata             &$  5.3_{- 0.3}^{+  0.2}$&Silver \\
203.3& 177.41123& 22.38846& \nodata &$4.8^{+0.2}_{-0.3}$ &\nodata             &$ 15_{- 3}^{+  5}$&Silver \\\cline{1-12}
204.1& 177.40961& 22.38666& \nodata &$6.1^{+2.0}_{-4.1}$ &$6.47_{-5.35}^{+0.18}$&$  8_{- 2}^{+  5}$&Silver \\
204.2& 177.40668& 22.38432& \nodata &$6.2^{+0.2}_{-4.8}$ &\nodata             &$  5.5 \pm 0.4$&Silver \\
204.3& 177.41208& 22.38905& \nodata &$6.8 \pm 0.3$       &\nodata             &$ 13 \pm 1$&Silver \\
\enddata
\tablecomments{All multiple images used in the gravitational mass model, with
  magnifications produced by the model, excluding the multiple images
  associated with SN Refsdal and its host galaxy (see \ref{tab:knots}).
  System 5 had only one spectroscopically-confirmed image; in the model, 
  the entire system is assumed to be at this redshift. System 13 had two 
  slightly different spectroscopic redshifts; the average of the redshifts 
  was used in the modeling. For original identification references, see 
  \citet{treu15}. Spectroscopic redshifts are obtained using data obtained 
  from the GLASS program and the Refsdal Follow-up Campaign. Photometric 
  redshifts (Peak P($z$)) are for individual images rather than combined 
  system redshifts. Combined system redshifts for photometric 
  systems are determined by combining the individual P($z$) using a Bayesian 
  combination procedure; the peak value for the resultant combined P($z$) 
  is considered to be the system redshift. In the System Redshift column 
  we have also included effective $1\sigma$ uncertainties for photometric 
  systems. These uncertainties are the lower and upper bound values which 
  yield $34\%$ on each side of the median combined P($z$) when integrating 
  under the P($z$) curve. The uncertainties are much larger than typical 
  photometric redshift uncertainties, given that we account for the 
  possibility of catastrophic errors by combining the individual system 
  P($z$) curves with a uniform PDF.}
\end{deluxetable*}

\clearpage
\LongTables
\begin{deluxetable*}{llllll}
\tablecolumns{6}
\tablewidth{0pc}
\tablecaption{Multiple Images of Star-Forming Knots in the SN Refsdal Host Galaxy \label{tab:knots} }
\tablehead{        
\colhead{ID} & 
\colhead{$\alpha$ (J2000)} & 
\colhead{$\delta$ (J2000)} & 
\colhead{Magnification}
}
\startdata
1.1.1&  177.39702& 22.39600& $  12 \pm 1$\\
1.1.2&  177.39942& 22.39743& $   7.3 \pm 0.5$\\
1.1.3&  177.40341& 22.40244& $   3.90 \pm 0.04$\\
1.1.5&  177.39986& 22.39713& $  12 \pm 2$\\\cline{1-6}
1.2.1&  177.39661& 22.39630& $  12.8 \pm 0.7$\\
1.2.2&  177.39899& 22.39786& $  23_{-  8}^{+ 16}$\\
1.2.3&  177.40303& 22.40268& $   3.96_{-  0.04}^{+  0.02}$\\
1.2.4&  177.39777& 22.39878& $  12.0_{-  0.6}^{+  1.1}$\\
1.2.6&  177.39867& 22.39824& $   2.8_{-  0.2}^{+  0.3}$\\\cline{1-6}
1.3.1&  177.39687& 22.39621& $  13.1_{-  0.9}^{+  0.6}$\\
1.3.2&  177.39917& 22.39760& $  13_{-  1}^{+  2}$\\
1.3.3&  177.40328& 22.40259& $   3.89 \pm 0.03$\\\cline{1-6}
1.4.1&  177.39702& 22.39621& $  13_{-  1}^{+  2}$\\
1.4.2&  177.39923& 22.39748& $  12_{-  1}^{+  2}$\\
1.4.3&  177.40339& 22.40255& $   3.88 \pm 0.04$\\\cline{1-6}
1.5.1&  177.39726& 22.39620& $  13 \pm 1$\\
1.5.2&  177.39933& 22.39730& $  12_{-  1}^{+  2}$\\
1.5.3&  177.40356& 22.40252& $   3.83_{-  0.07}^{+  0.04}$\\\cline{1-6}
1.6.1&  177.39737& 22.39616& $  15 \pm 1$\\
1.6.2&  177.39945& 22.39723& $   8.8_{-  0.7}^{+  0.9}$\\
1.6.3&  177.40360& 22.40248& $   3.83_{-  0.07}^{+  0.04}$\\\cline{1-6}
1.7.1&  177.39757& 22.39611& $  18 \pm 2$\\
1.7.2&  177.39974& 22.39693& $  12 \pm 2$\\
1.7.3&  177.40370& 22.40240& $   3.84_{-  0.08}^{+  0.04}$\\\cline{1-6}
1.8.1&  177.39795& 22.39601& $  27_{-  3}^{+  5}$\\
1.8.2&  177.39981& 22.39675& $   9.4 \pm 0.5$\\
1.8.3&  177.40380& 22.40231& $   3.74_{-  0.10}^{+  0.05}$\\\cline{1-6}
1.9.1&  177.39803& 22.39593& $  27_{-  3}^{+  5}$\\
1.9.2&  177.39973& 22.39698& $  12 \pm 2$\\
1.9.3&  177.40377& 22.40225& $   3.88_{-  0.08}^{+  0.06}$\\\cline{1-6}
1.10.1& 177.39809& 22.39585& $  46_{- 12}^{+  18}$\\
1.10.2& 177.39997& 22.39670& $   6.7_{-  0.4}^{+  0.5}$\\
1.10.3& 177.40380& 22.40218& $   3.88_{-  0.08}^{+  0.06}$\\\cline{1-6}
1.11.2& 177.40010& 22.39666& $   4.6_{-  0.4}^{+  0.5}$\\
1.11.3& 177.40377& 22.40204& $   3.94_{-  0.07}^{+  0.04}$\\\cline{1-6}
1.12.1& 177.39716& 22.39521& $   7.9_{-  0.5}^{+  0.6}$\\
1.12.2& 177.40032& 22.39692& $   6.1_{-  0.3}^{+  0.4}$\\
1.12.3& 177.40360& 22.40187& $   4.07_{-  0.05}^{+  0.04}$\\\cline{1-6}
1.13.1& 177.39697& 22.39663& $  17.7_{-  1.1}^{+  0.8}$\\
1.13.2& 177.39882& 22.39771& $  67_{-206}^{+140}$\\
1.13.3& 177.40329& 22.40282& $   3.90_{-  0.05}^{+  0.03}$\\
1.13.4& 177.39791& 22.39843& $  22_{-  3}^{+  7}$\\\cline{1-6}
1.14.1& 177.39712& 22.39672& $  20_{-  2}^{+  1}$\\
1.14.2& 177.39878& 22.39763& $  27_{-  7}^{+ 12}$\\
1.14.3& 177.40338& 22.40287& $   3.63_{-  0.08}^{+  0.02}$\\
1.14.4& 177.39810& 22.39825& $  48_{- 22}^{+114}$\\\cline{1-6}
1.15.1& 177.39717& 22.39650& $  20 \pm 2$\\
1.15.2& 177.39894& 22.39751& $  25_{-  5}^{+  8}$\\
1.15.3& 177.40344& 22.40275& $   3.81_{-  0.05}^{+  0.03}$\\\cline{1-6}
1.16.1& 177.39745& 22.39640& $  19 \pm 2$\\
1.16.2& 177.39915& 22.39722& $  15 \pm 1$\\
1.16.3& 177.40360& 22.40265& $   3.67_{-  0.10}^{+  0.02}$\\\cline{1-6}
1.17.1& 177.39815& 22.39634& $  26_{-  6}^{+  7}$\\
1.17.2& 177.39927& 22.39683& $  22_{-  2}^{+  3}$\\
1.17.3& 177.40384& 22.40256& $   3.67_{-  0.10}^{+  0.04}$\\\cline{1-6}
1.18.1& 177.39850& 22.39610& $ 105_{- 44}^{+102}$\\
1.18.2& 177.39947& 22.39659& $  14 \pm 1$\\
1.18.3& 177.40394& 22.40240& $   3.68_{-  0.12}^{+  0.05}$\\\cline{1-6}
1.19.1& 177.39689& 22.39576& $  10.2 \pm 0.6$\\
1.19.2& 177.39954& 22.39748& $   5.6_{-  0.4}^{+  0.3}$\\
1.19.3& 177.40337& 22.40229& $   4.1 \pm 0.2$\\
1.19.5& 177.39997& 22.39710& $  16_{-  7}^{+ 33}$\\\cline{1-6}
1.20.1& 177.39708& 22.39572& $   9.2 \pm 0.6$\\
1.20.2& 177.39963& 22.39736& $   6.2_{-  0.4}^{+  0.7}$\\
1.20.3& 177.40353& 22.40223& $   3.92_{-  0.08}^{+  0.03}$\\
1.20.5& 177.40000& 22.39698& $  14_{-  7}^{+  26}$\\\cline{1-6}
1.21.1& 177.39694& 22.39540& $   8.0_{-  0.7}^{+  0.8}$\\
1.21.3& 177.40341& 22.40200& $   4.12_{-  0.02}^{+  0.03}$\\
1.21.5& 177.40018& 22.39704& $  25_{- 49}^{+ 50}$\\\cline{1-6}
1.22.1& 177.39677& 22.39548& $   8.6_{-  0.8}^{+  0.6}$\\
1.22.2& 177.39968& 22.39749& $   5.6_{-  0.4}^{+  0.3}$\\
1.22.3& 177.40328& 22.40209& $   4.17_{-  0.04}^{+  0.06}$\\
1.22.5& 177.40008& 22.39713& $  44_{-105}^{+ 76}$\\\cline{1-6}
1.23.1& 177.39672& 22.39538& $   7.8_{-  0.5}^{+  0.7}$\\
1.23.2& 177.39977& 22.39749& $   5.3 \pm 0.3$\\
1.23.3& 177.40324& 22.40201& $   4.17_{-  0.04}^{+  0.06}$\\
1.23.5& 177.40013& 22.39720& $  44_{-105}^{+ 76}$\\\cline{1-6}
1.24.1& 177.39650& 22.39558& $   8.8_{-  0.9}^{+  0.6}$\\
1.24.2& 177.39953& 22.39775& $   6.5_{-  0.3}^{+  0.7}$\\
1.24.3& 177.40301& 22.40220& $   4.19_{-  0.03}^{+  0.04}$\\\cline{1-6}
1.25.1& 177.39657& 22.39593& $  10.9_{-  0.8}^{+  0.7}$\\
1.25.3& 177.40304& 22.40245& $   4.00 \pm 0.03$\\\cline{1-6}
1.26.1& 177.39633& 22.39601& $   8.8_{-  0.8}^{+  0.7}$\\
1.26.3& 177.40283& 22.40260& $   4.10_{-  0.03}^{+  0.02}$\\\cline{1-6}
1.27.1& 177.39831& 22.39628& $ 138_{-478}^{+335}$\\
1.27.2& 177.39933& 22.39672& $  20 \pm 2$\\\cline{1-6}
1.28.1& 177.39860& 22.39616& $ 104_{- 44}^{+102}$\\
1.28.2& 177.39942& 22.39655& $  18_{-  1}^{+  2}$\\\cline{1-6}
1.29.1& 177.39858& 22.39586& $  19_{-  2}^{+  3}$\\
1.29.2& 177.39976& 22.39649& $  12.6_{-  0.5}^{+  0.7}$\\\cline{1-6}
1.30.1& 177.39817& 22.39546& $  30_{-  5}^{+ 13}$\\
1.30.2& 177.39801& 22.39523& $  24_{-  3}^{+  4}$\\
1.30.3& 177.39730& 22.39536& $   8.7_{-  1.0}^{+  0.7}$\\
1.30.4& 177.39788& 22.39572& $  41_{- 12}^{+ 17}$\\\cline{1-6}
S1    & 177.39823& 22.39563& $  30_{-  4}^{+  6}$\\
S2    & 177.39772& 22.39578& $  22 \pm 3$\\
S3    & 177.39737& 22.39553& $  11 \pm 1$\\
S4    & 177.39781& 22.39518& $  15 \pm 2$\\
SX    & 177.40024& 22.39681& $   6.30_{-  0.4}^{+  0.5}$\\
SY    & 177.4038 & 22.40214& $   3.88_{-  0.08}^{+  0.06}$\\
\enddata
\tablecomments{All star-forming knots in the host galaxy of SN Refsdal
  ($z_s=1.49$), as well as all images of the supernova, with magnifications
  produced by the model. Objects S1-S4, SX, and SY are images of SN Refsdal;
  they are also used as model inputs. For original identification references,
  see \citet{treu15}. Magnification uncertainties allowing for negative values
  indicate that crossing the critical curve lies in the 68\% confidence
  interval of magnification values.}
\end{deluxetable*}

\section{Mass Modeling}
\subsection{Summary of the Modeling Procedure}
In order to model the mass distribution of MACS1149, we use the Strong and 
Weak Lensing United (SWUnited) code developed by \citet{bradac09, bradac05}, 
which has previously been used to model other HFF clusters \citep{hoag16, 
wang16}.\footnote{\url{https://archive.stsci.edu/prepds/frontier/lensmodels}} 
This software extends weak lensing reconstruction to critical regions, thus
using information from both strongly-lensed multiple images and a catalog of
weakly-lensed background galaxies to break the mass-sheet degeneracy. It also
makes no explicit assumption that light traces the underlying dark matter
distribution, which is ideal for comparing the distributions of stellar and
dark matter components.

After beginning with an initial mass model, the procedure minimizes a penalty
function, $\chi^2$, which takes into account multiple image positions
($\chi^2_{SL}$), weak lensing galaxy ellipticities ($\chi^2_{WL}$), and a
regularization term ($R$) that promotes solutions with few unphysical small-scale
fluctuations:
\begin{equation}
\chi^2(\psi_k) = \chi^2_{SL}(\psi_k) + \chi^2_{WL}(\psi_k) + \eta R(\psi_k),
\end{equation}
where $\eta$ is the regularization parameter (we used a value of 
$\eta$=0.75, though the exact choice of $\eta$ is somewhat arbitrary). 
With each iteration, the procedure minimizes 
the penalty function for a trial solution and then increases the number of 
grid points. Thus, each trial solution produced is more detailed than the 
previous solution, until the code converges on a solution that has errors
consistent with the errors of the original constraints. Further details, 
including discussion of how the penalty function is defined and minimized,
are given by \citet{bradac05}.

The mass map has a non-uniform resolution because in regions with high 
signal-to-noise ratio (i.e., regions which are well-constrained with multiple 
images), we increase the resolution by dividing the model's effective 
pixel scale by 2 with each iteration; we call this process refinement. We 
implement five levels of refinement: the lowest level for low model $S/N$ (does 
not divide pixels and maintains an effective pixel scale of $4\farcs8$; 
applied in regions of the map with no multiple images constraining the
model, such as the outskirts), the highest level for high $S/N$ (divides
each pixel into $16\times16$ pixels for an effective pixel scale of
$0\farcs3$; applied in regions with excellent constraints such as the regions
in which star-forming regions of the SN Refsdal host galaxy are located), and
three levels in between to help avoid artifacts from sudden change in
effective resolution. Regions with multiple images not associated with the
host galaxy of SN Refsdal are refined to the third-highest level. Intermediate
levels are also used around areas of high mass density, such as the two large
galaxies to the southwest of the BCG and the association of galaxies to the
northeast of the BCG, to improve the model's flexibility in these regions.

\subsection{Modeling Inputs}
For spectroscopically-confirmed multiple images the spectroscopic redshift
is used as a constraint to the model. For the multiple images without
spectroscopic redshifts, we use photometric redshifts. Photometric redshifts 
have larger uncertainties, and they do not always agree exactly within the 
same system of multiple images. To determine one input redshift to our model 
for each multiply-imaged system, we combine the P($z$)'s within the system 
using the hierarchical Bayesian method described in \citet{wang16}. We use
the peak redshift of the combined P($z$) as the input redshift to the model. 

Reconstruction was performed on the HST ACS field of view 
($202\farcs\times202\farcs$). We assume a $0\farcs5$ positional uncertainty for 
strongly-lensed images when computing the $\chi^2$. All redshifts are listed 
in Tables 1 and 2.

\subsection{Modeling Results}
\begin{figure}
  \includegraphics[width=0.49\textwidth]{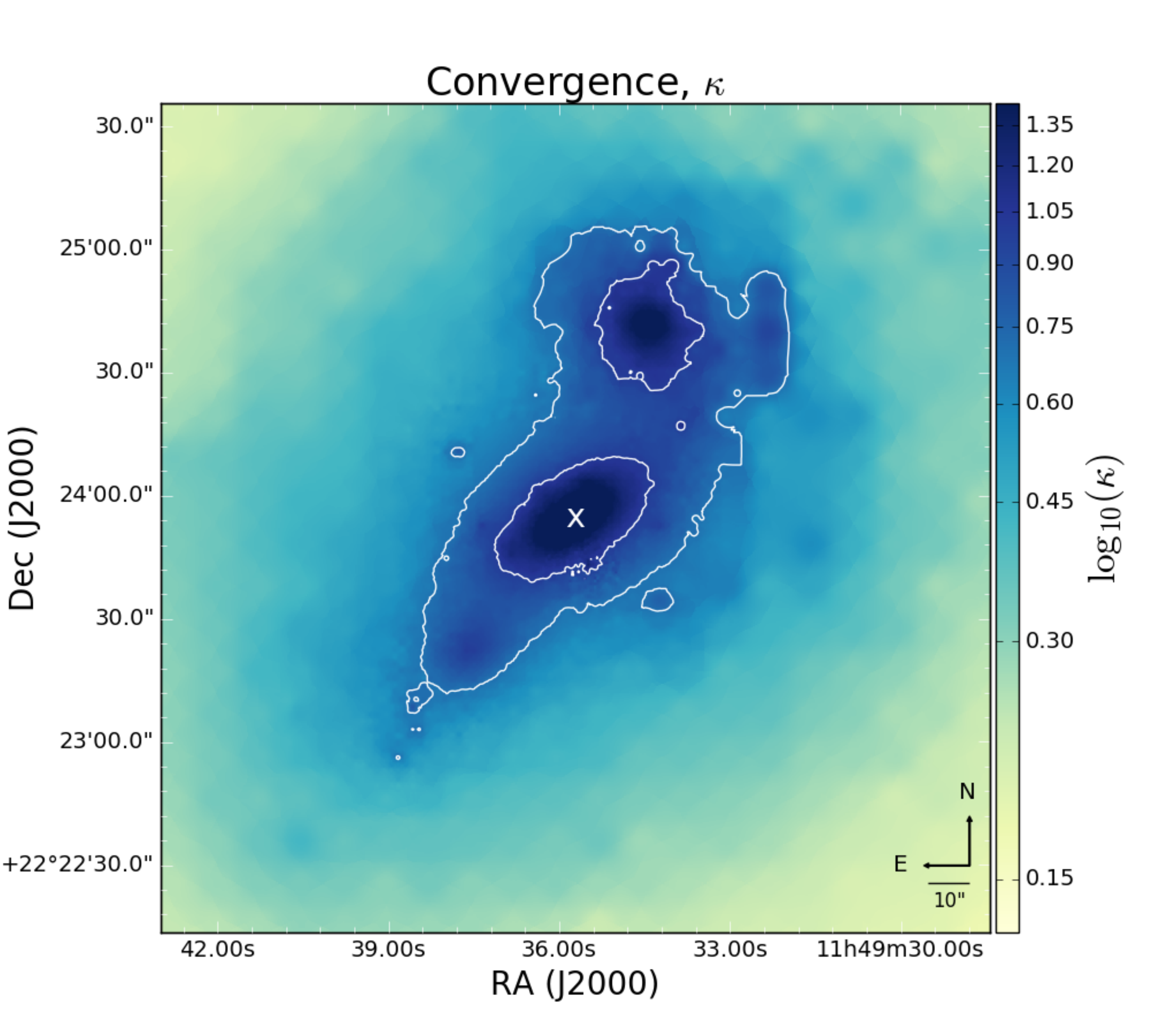}
  \includegraphics[width=0.49\textwidth]{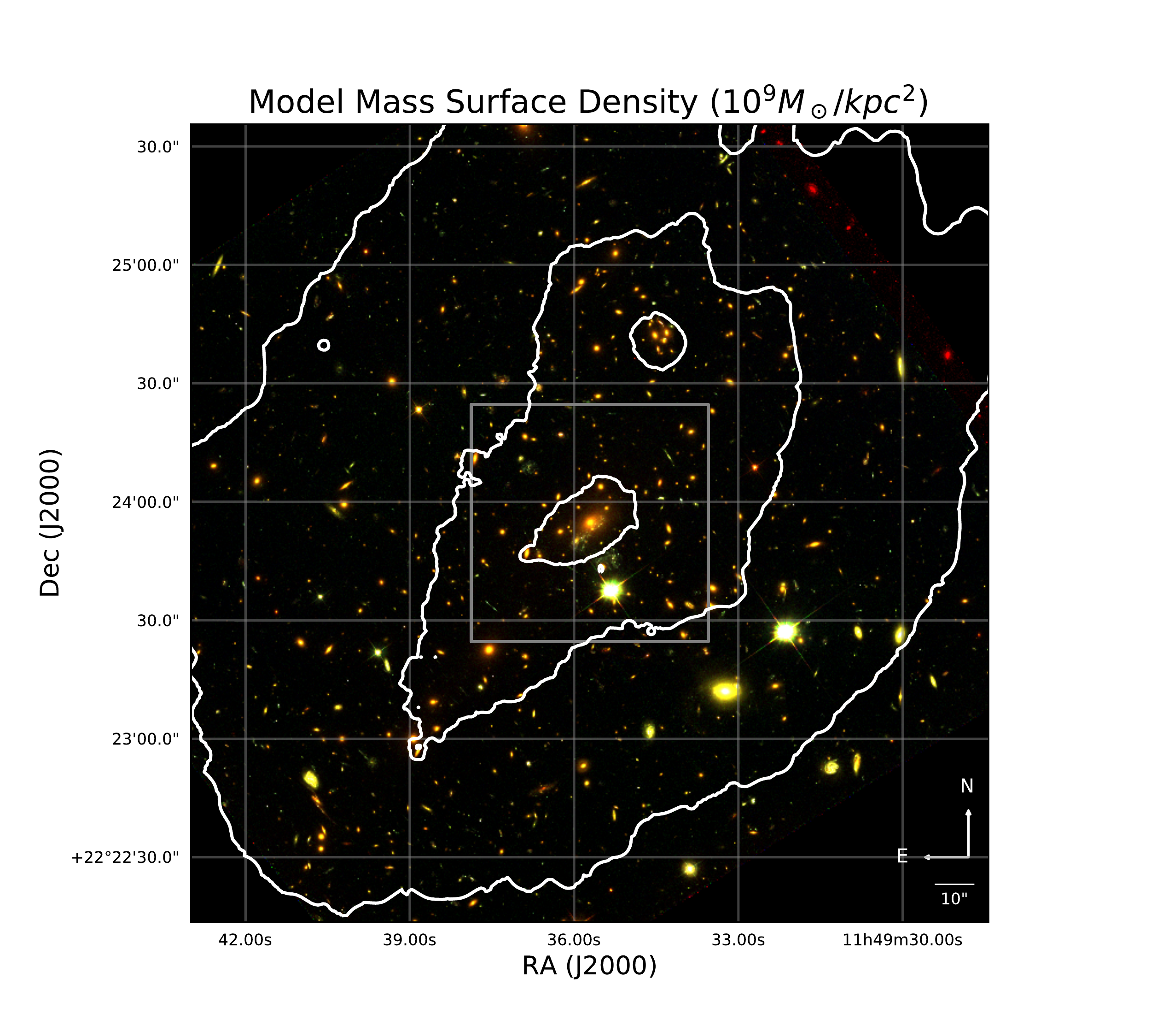}
  \caption{\label{fig:kappa} \textbf{Top:} Convergence ($\kappa$) map for 
  $z_s=9$ over the HST ACS field of view ($202\times202\farcs$). The
    convergence reveals an elliptical total mass distribution with two
    major peaks, one occurring at the BCG and one occurring at a group of
    galaxies to the northwest of the BCG. The differences in effective
    resolution across regions are evident in this image --- regions toward the
    outskirts of the modeling region are more pixellated than those near the
    BCG, as discussed in section 3.2. Note that this image employs a
    logarithmic stretch, and colorbar values show $\mathrm{log}_{10}(\kappa$).
    Contours shown correspond to $\kappa = 5$ and $\kappa = 10$. The cluster 
    BCG is labelled with a white x. \textbf{Bottom:} Mass surface
    density contours for our mass model over the HST ACS field of view. Contours
    correspond to mass surface density values of 0.5, 1.0, and 2.0 
    $\times 10^9 M_{\odot}/\textrm{kpc}^2$.}
\end{figure}

\begin{figure*}
\centering
\includegraphics[width=\textwidth]{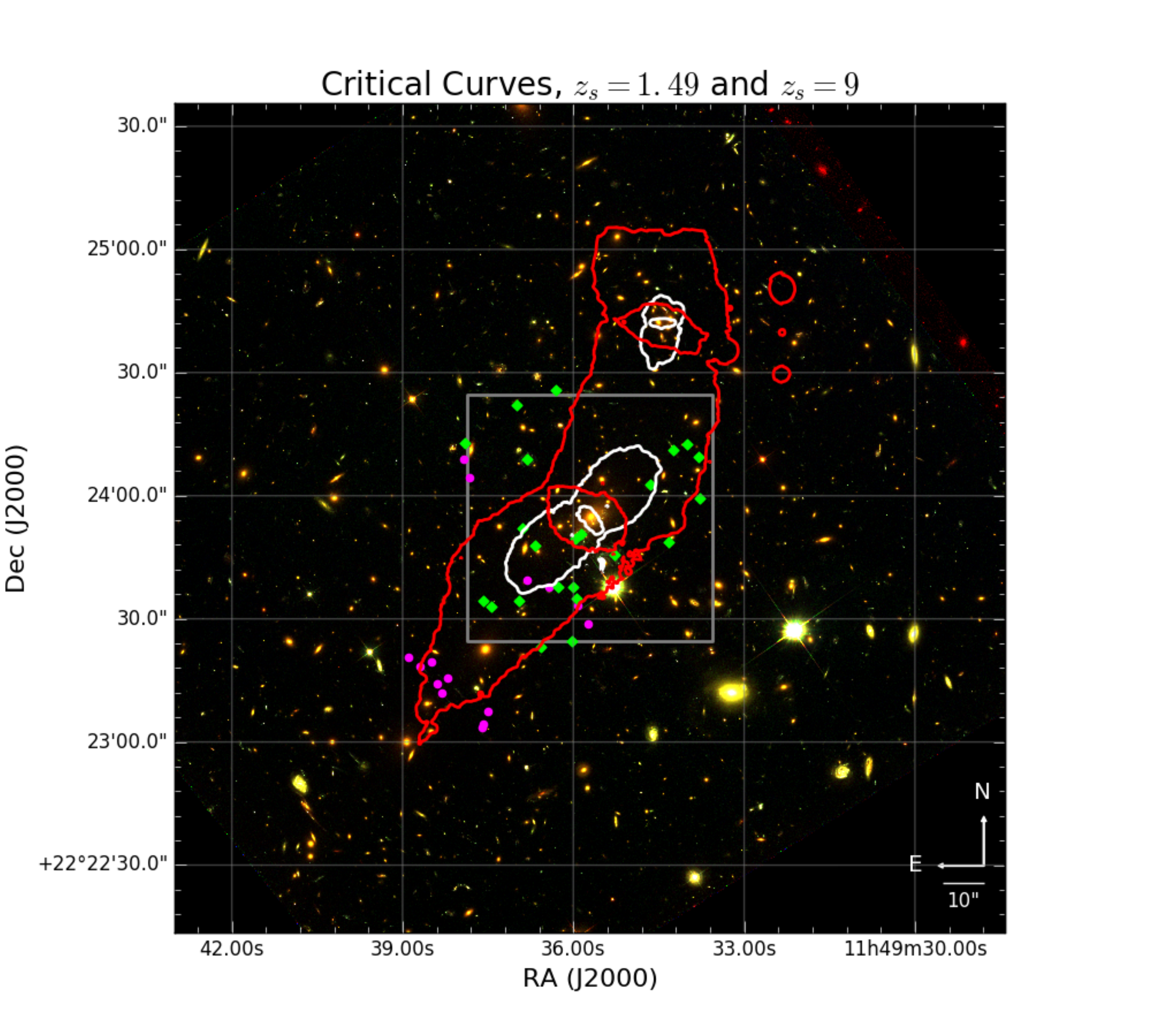}
\caption{\label{fig:critcurves} Critical curves for redshifts $z_s=1.49$
    (white), the redshift of SN Refsdal, and $z_s=9$ (red), over the HST ACS
    field of view ($202\times202\farcs$). Multiple image 
    locations are shown; green points are spectroscopically-confirmed images 
    and magenta points correspond to images with photometric redshifts only. 
    The multiple images are listed in Table 1. A zoomed-in image comparing 
    this model's critical curves at $z_s=1.49$ to other modelers' critical 
    curves is shown in Figure \ref{fig:allcritcurves}; its spatial extent 
    is given by the gray box.}
\end{figure*}

A map of the convergence for source redshift $z_s=9$ is shown in Figure 
\ref{fig:kappa}. There are two predominant peaks in the convergence map ---
one near the BCG, and a smaller peak to the northwest that coincides with an
overdensity of cluster galaxies. To the southeast of the BCG there is a small
peak corresponding to a large cluster member. We find $83.4\%^{+0.2}_{-0.3}$ of
the image plane area modeled contains magnification values between $\mu=1$
and $\mu=4$, with magnifications reaching $\mu = 10-20$ a few arcseconds from
the critical curves (Figure \ref{fig:critcurves}), and a median absolute
magnification of $1.88\pm0.01$ over the field of view of the Hubble Wide Field
Camera 3 (WFC3).

\begin{figure*}
\centering
\includegraphics[width=\textwidth]{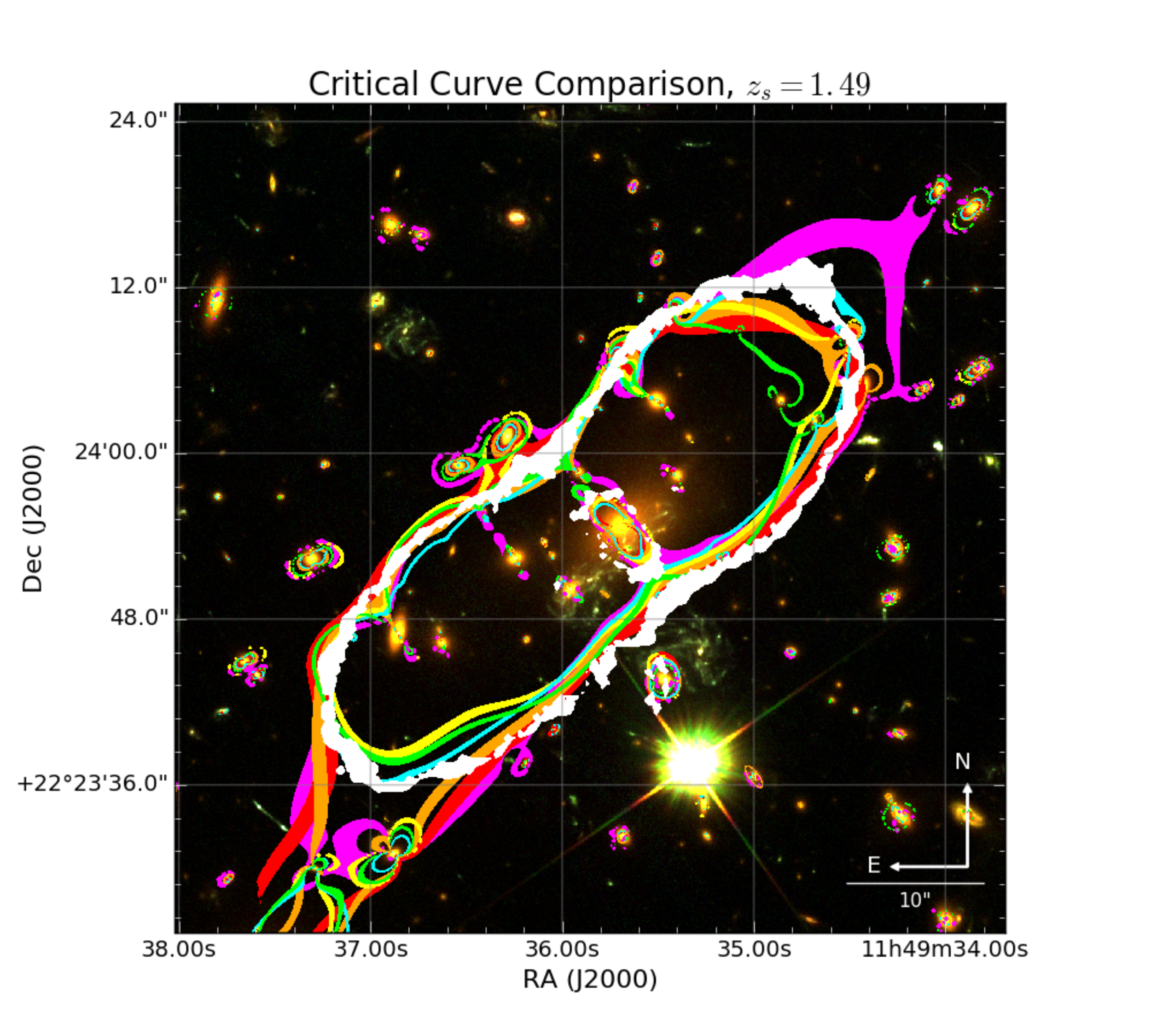}
  \caption{\label{fig:allcritcurves} A comparison of our critical curves at
    $z_s=1.49$ (white) with those of \citet[][magenta]{diego15},
    \citet[][red]{grillo15a}, \citet[][orange, all images model]{kawamata15},
    \citet[][yellow]{jauz16}, \citet[][cyan, gold-only model]{sharon15},
    and Zitrin et al. (in prep, green). The general shape of these modelers'
    critical curves is in close agreement with those produced by our model.
    The critical curves shown are created by filling in regions of high
    magnification; the thickness of the curves depends on each model's
    resolution. The field of view for this image is $60\farcs\times60\farcs$,
    and is indicated by the gray square in Figure \ref{fig:critcurves}.}
\end{figure*}

In our model, we use all gold images and most silver images, differing
from those teams who created gold-only models \citep[][Zitrin et al, in
prep]{grillo15a, kawamata15, sharon15} or models that used all gold and silver
images \citep[][note two teams created a model for each set of multiple
images]{diego15, kawamata15, sharon15}. We also reconstruct our mass model
on an adaptive grid using weak lensing as input, which prominently differs
from all modeling techniques discussed in \citet{treu16}. Despite these
distinctive inputs and modeling procedure, our model's critical curves
closely match those of the other groups in the \citet{treu16} collaboration
(Figure \ref{fig:allcritcurves}, reflecting the strength of the constraints
available to model this cluster.

We note that the constraints to the model in the central regions of the
cluster are strong, but there are few constraints in the northwest region,
the location of the object MACS1149-JD \citep{zheng12}. An analysis of this
object, including our model's estimates for its magnification, will be
presented in a forthcoming paper (Hoag et al. 2017, in prep). 

The overall root-mean-square (RMS) offset for the images, including both 
strongly-lensed background galaxy images and star-forming regions of the host 
galaxy for SN Refsdal, is $1''$, competitive with other models for the cluster 
\citep{treu16}. While it would be possible for our methodology to yield an 
RMS of $0''$, we tune the regularization parameter $\eta$ to prevent this 
outcome so as to avoid overfitting to noise. 

At redshift $z_s=1.49$ (the redshift of SN Refsdal) we are able to closely
reproduce the positions of the knots of the spiral host galaxy of SN Refsdal,
although unlike parametric models, the effective resolution of our model
($0\farcs3$/pixel) is too low to resolve the mass of the
host galaxy in sufficient detail. Thus, we are unable to accurately 
determine the critical curves around the galaxy-scale lensing of SN 
Refsdal (images S1, S2, S3, and S4, see Table 2), or to make reliable 
predictions of the time delay between SN Refsdal images. Despite these 
limitations related to our effective resolution that prohibit us from making 
estimates based on fine-scale details of the cluster model, we have confidence 
in the large-scale shape of the model (Figure \ref{fig:allcritcurves}). 

As a response to an STScI call to model the HFF clusters before and
after the arrival of HFF data, this code was previously used to model MACS1149
using CLASH data. We compare our current (v3) model with the previous (v1)
model using CLASH data. Note that some groups presented models in between the
two official calls, which they denoted as v2; we did not create a v2
model.\footnote{\label{mast} \url{https://archive.stsci.edu/pub/hlsp/frontier/macs1149/models/}} In the
previous model, we used only 38 multiple images (14 systems), two of which
(systems 8 and 12) have since been proven with spectroscopy to be spurious
identifications. With updated imaging and spectroscopic data, the current
version of the lensing model (v3) is substantially different from the previous
model, much more similar to the other models (Figure \ref{fig:allcritcurves})
produced with the same post-HFF data.  All maps of convergence, shear, and
magnification, from this modeling team and other teams, are included in the
Mikulski Archive for Space Telescopes (MAST, see footnote \ref{mast}).

To determine the error in our model associated with photometric redshift
uncertainty or incorrect system labels, we resample from the photometric 
redshift probability density functions $P_i(z_{\rm{Bayes}})$ produced by EAZY. 
We allow resampled photometric redshifts to have values 
$z_{\rm{phot}} < z_{\rm{cluster}}$, to account for the possibility of catastrophic 
errors. Although system redshift estimates may vary during this resampling, 
images belonging to a single system are always modeled as having a single 
redshift. After resampling the photometric systems, we run the modeling code 
again, and find that varying redshift estimates for photometric systems in the 
model produces errors within the range we have quoted.

We also determine the uncertainty associated with choice of weak lensing 
galaxies and SN Refsdal host galaxy star-forming regions using a 
bootstrap-resampling procedure that randomly weights the weak lensing and
Refsdal system inputs. After this bootstrapping, we remodel the cluster with
the bootstrapped values, and find that errors associated with our catalogs for
weak lensing and SN Refsdal host galaxy star-forming regions are subdominant
to the error associated with photometric redshifts. Nevertheless, we propagate
all of these sources of error into the final errors for all lensed quantities.
Additionally, we did vary the initial model, and found the results 
converge within error bars given by bootstrap estimates.

We assume any error in spectroscopic redshifts is small in comparison to
modeling uncertainties, and we do not fully account for all 
possible sources of systematic error, such as choice of initial mass model, 
choice of refinement grid, spectroscopic redshift error, choice of 
regularization parameter, or spectroscopic image identification error. 

\section{Stellar Mass to Total Mass Ratio}
\subsection{Details of the Stellar Mass Density Map}
We estimate the stellar mass density of MACS1149 with measurements of the
approximate rest-frame K-band flux from the cluster members and a 
mass-to-light ratio, in the same manner as \citet{hoag16}. We summarize this
procedure in the following section.

\begin{figure}
  \includegraphics[width=0.49\textwidth]{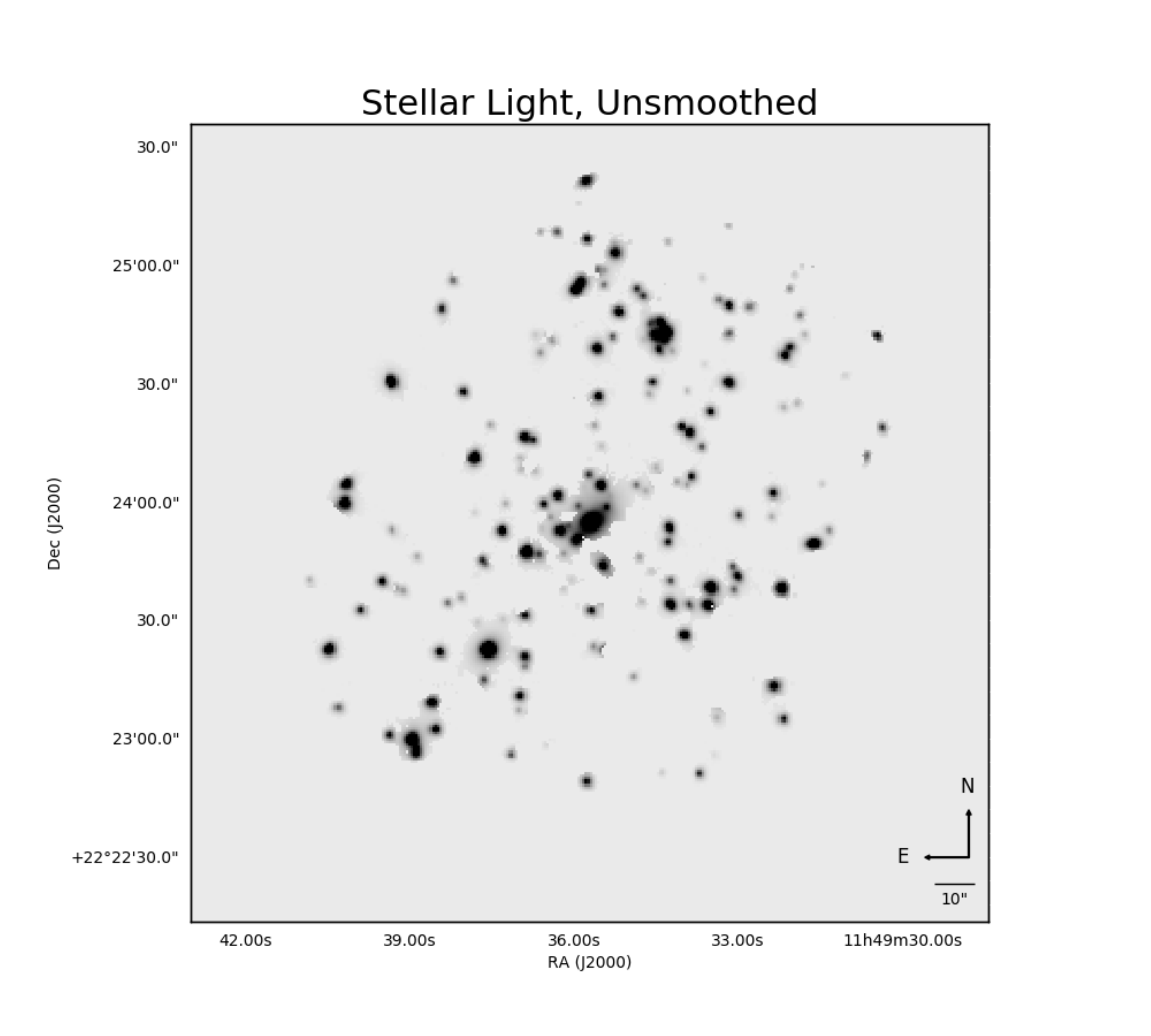}
  \caption{\label{fig:clmembers} Unsmoothed light map of the cluster
  members used in the creation of the $f_{\star}$ map. We use a cluster
  membership procedure based on astrometric matching and redshift cuts to
  create a cluster member catalog of 591 cluster galaxies. This light
  map is subsequently smoothed and compared to the mass model map to 
  produce a map of the stellar to total mass ratio ($f_{\star}$, 
  Figure \ref{fig:fstar})}
\end{figure}

We determine cluster membership using the catalog produced by
\citet{morishita16}. We use 123 spectroscopic cluster members obtained from 
the public catalog of \citet{ebeling14}, and 530 cluster members chosen using
the photometric redshifts with informative prior presented by
\citet{morishita16}, for a total of 591 cluster galaxies. Cluster member 
galaxies must obey the selection criterion
$\delta z \equiv |z-z_{cls}|/(1+z_{cls}) < \delta z_{cut}$, where $z_{cls}$ is the
redshift of the cluster and $z$ is the redshift of each object. The cutoff
redshift range $\delta z_{cut} = 0.0084$ for those galaxies chosen from the
\citet{ebeling14} spectroscopic catalog, and $\delta z_{cut} = 0.0219$ for
galaxies chosen by photometric redshift. (The average photo-z error is
$\sigma_z = 0.0073$, so this is a $3\sigma$ cutoff value.) Our catalog, which
primarily consists of photometrically-determined cluster members, is much 
larger than the primarily spectroscopically-determined cluster catalog of
\citet{grillo15a}, ensuring that we incorporate all cluster light into our
estimates for stellar mass surface density, down to a completeness limit
of $\textrm{log} M_{\star}/M_{\odot} = 7.8$. An unsmoothed light map of 
the catalog is shown in Figure \ref{fig:clmembers}.

To measure the flux from the cluster members, a mask is created to
eliminate all light in \emph{Spitzer}/IRAC channel 1 that is not
associated with a cluster member, and then we convolve with a one-pixel
Gaussian kernel to smooth over the mask boundary. This rest-frame
luminosity map is converted into a stellar mass surface density map
using the mass-to-light ratio determined by \citet{bell03} using their
''diet-Salpeter'' initial mass function (IMF). We use the mass-to-light ratio 
$M_{\star}/L = (0.83 \pm 0.26) M_{\odot}/L_{\odot}$, which corresponds 
to the $10<$ log$(M_{\star}h^2)<10.5$ stellar mass bin of the analysis by 
\citet{bell03}. Note that given the substantial differences in $M_{\star}/L$
ratio between IMFs, our choosing an IMF introduces the largest source 
of systematic uncertainty, as large as $\delta f_{\star} = 0.005$. 
While this uncertainty is large, we note that this approach to 
choosing the IMF is standard in SED fitting \citep{bell03, ilbert10, ilbert13, muzzin13, tomczak14, tomczak17}. It is also possible the IMF may differ
across galaxies in the cluster. While we do not account for this source of
error, we note that the massive, early-type galaxies in the cluster center
are more likely to exhibit an IMF closer to Salpeter \citep{auger10}. As a
result, we anticipate that the central peak in the stellar mass density map 
would be enhanced if we were to model dominant galaxies with a heavier IMF.

Further, we adopt conservatively-large error bars on this mass-to-light
ratio, using the spread of the $M_{\star}/L$ distribution as our statistical
uncertainty on the $M_{\star}/L$ ratio rather than using the width of the most
likely mean $M_{\star}/L$ bin \citep{bell03}. This choice accounts for
possible correlation between the $M_{\star}/L$ ratios of galaxies in a single
cluster, for a fixed IMF. Typically, when finding an average parameter value
in a binned distribution of parameter means, it is appropriate to choose the
width of the bin as the value for the statistical uncertainty. However, if the
$M_{\star}/L$ ratios of galaxies in MACS1149 cluster were correlated, this
approach would be insufficient to capture the uncertainty of the estimate,
since galaxy $M_{\star}/L$ ratios could deviate from the mean in a related
fashion. Thus, we must account for the spread of all possible $M_{\star}/L$
ratios to make a conservative estimate on the $M_{\star}/L$ ratios for these
galaxies. In doing so, we incorporate a $\sim 30\%$ statistical uncertainty on
our $M_{\star}/L$ ratio into our estimates, an uncertainty which dominates the
statistical error in the stellar mass density model. This $\sim 30\%$
uncertainty accounts for possible variation of $M_{\star}/L$ of individual cluster
members; this error is in addition to the systematic error associated with
the choice of IMF.

To account for the additional stellar mass associated with the intracluster
light (ICL), we increase the stellar mass surface densities across the map 
using the $8^{+12}_{-4}$\% ICL mass estimate for MACS1149 by 
\citet{morishita16b}. We propagate the errors in the ICL through to the
final stellar mass surface density map. However, the effect of the ICL is
still subdominant to both the statistical error on the $M_{\star}/L$ ratio and
the systematic error associated with choosing an IMF in determining the 
uncertainty of the stellar mass surface density map. We do not account 
for possible radial gradients in the ICL. \citet{morishita16b} have noted that
the ICL in MACS1149 is centrally concentrated with a shallow slope to $R<200$ kpc
from the cluster center. Thus, we expect any radial gradients in the ICL to 
enhance the central peak in the stellar mass density map.

We also acknowledge that our procedure for choosing cluster members 
introduces the possibility of impurities in the catalog construction. However, 
we do not expect a spatial variation in these impurities, and thus anticipate 
that they would not effect our overall conclusions beyond a small bias factor 
in the stellar to total mass ratio.

To determine statistical errors in the maps, we resample from a Gaussian
distribution for $M_{\star}/L$ with mean and standard deviation given by the
distribution in \citet{bell03}. We then use a resampling procedure to find
the variation in the total mass surface density maps, and multiply the
stellar mass surface density maps with an ICL resampled from the
distribution from \citet{morishita16b} to yield a set of final error maps.


\subsection{Details of the $f_{\star}$ Map}
In our modeling approach, we do not directly assume that stellar light traces
dark matter mass. The total mass distribution derived using our model is not
strictly coupled to the stellar mass distribution. Therefore, it is
informative to compare the spatial distribution of the ratio of stellar mass
to total mass, $f_{\star}$. We further define the average ratio of the stellar
mass to the total mass in an aperture as $\langle f_{\star} \rangle$.

To produce a projected $f_{\star}$ map, we match the mass resolution of the
stellar mass density map with that of our model-constructed total mass 
density map, which has non-uniform mass resolution due to our method of 
refinement. To match the stellar mass density resolution to that of the total 
mass density map, we create five stellar mass surface density maps of
different resolutions by convolving with five different Gaussian kernels of
width corresponding to each of our five different levels of refinement.
We use the best-fit kernel width of $0\farcm74$ for the level of lowest
refinement, as empirically determined by \citet{hoag16}. The map
corresponding to the second-lowest level of refinement is convolved with a
Gaussian kernel with half the width of the kernel for the lowest level of
refinement, and each subsequent level is convolved with a kernel smaller by
a factor of two than the previous level of refinement. We then add the maps,
weighting the pixels of each map by their locations on the refinement grid.

\begin{figure}
  \includegraphics[width=0.49\textwidth]{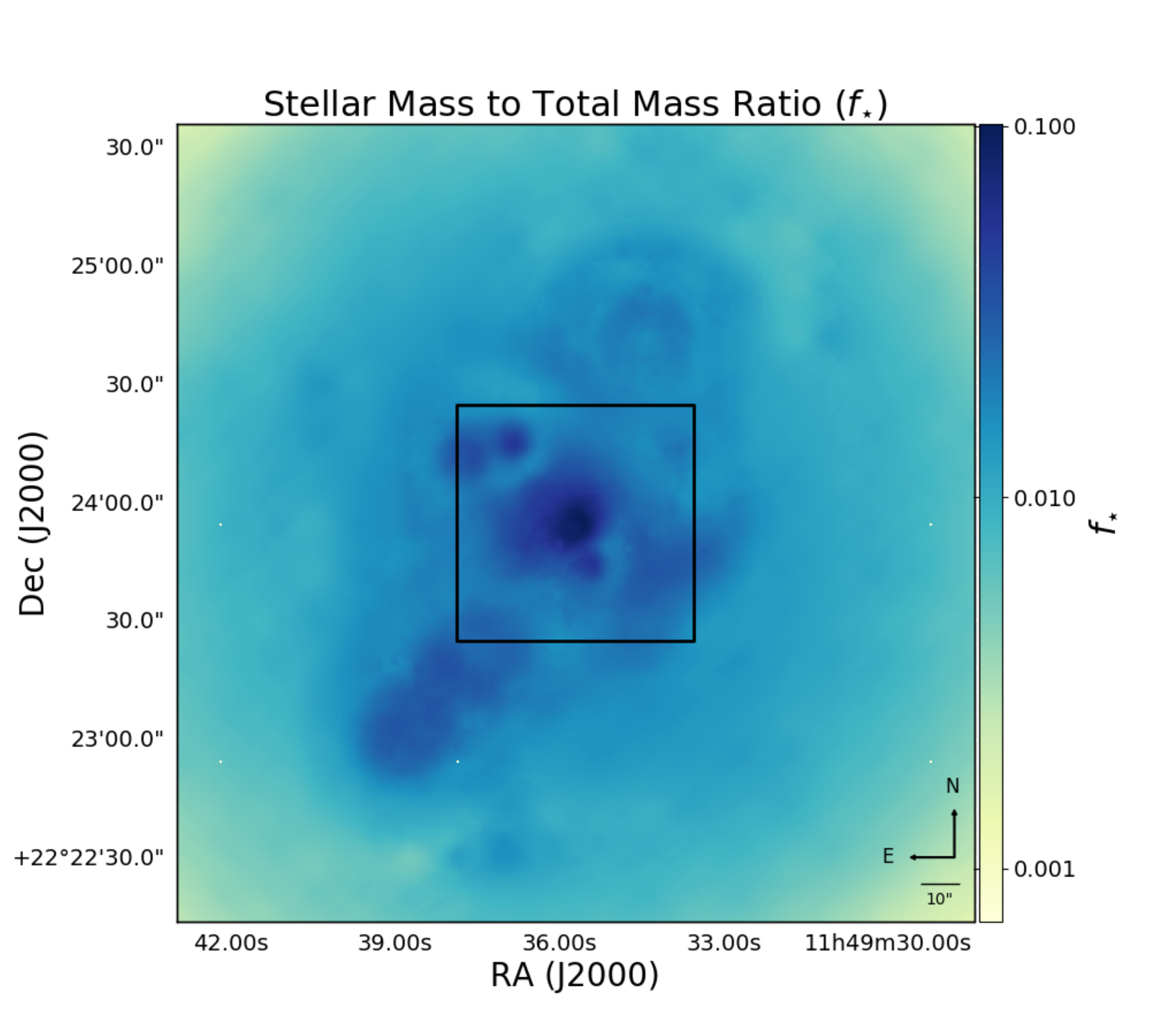}
  \includegraphics[width=0.49\textwidth]{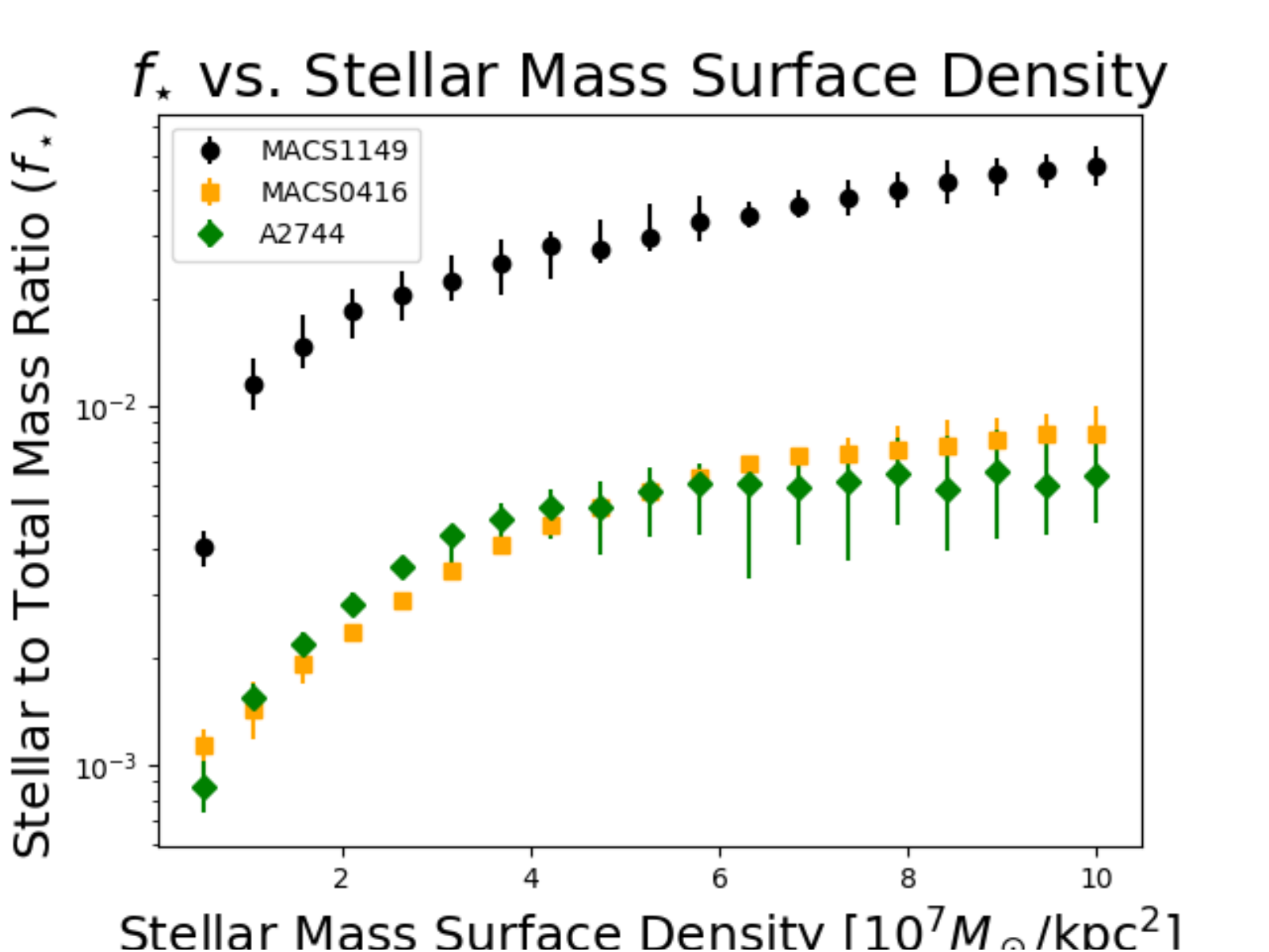}
  \caption{\label{fig:fstar} \textbf{Top: } Spatial distribution of projected 
    stellar mass to total mass ratio ($f_{\star}$). The center of the cluster
    has a value approaching $f_{\star}=5\%$, and the $f_{\star}$ value decreases
    with increasing distance from the mass concentrations in the cluster.
    Note that this image employs a logarithmic stretch, and colorbar values
    show $\mathrm{log}_{10}(f_{\star}$). The spatial extent of Figure 
    \ref{fig:allcritcurves} is given by the black box.
    \textbf{Bottom: } Stellar mass to total mass ratio within our
  field of view ($\langle f_{\star} \rangle$) vs. stellar mass surface density
  (in units of $10^{7} M_\odot/\rm{kpc}^2$), for the clusters MACS1149 (black 
  circles), MACS0416 \citep[orange squares,][]{hoag16} and A2744 \citep[green 
  diamonds,][]{wang16}. Given the complexity of these clusters' substructures 
  and ambiguities in precisely defining the cluster center, this depiction of 
  $\langle f_{\star} \rangle$ across varying physical scales is more reliable 
  than a plot of $\langle f_{\star} \rangle$ vs. aperture size. Almost 
  uniformly, in regions with higher stellar mass surface density there is a 
  higher stellar mass fraction, particularly near the cluster core. MACS0416 and A2744 in general have much lower values of $\langle f_{\star} \rangle$ at 
  these stellar mass surface densities. Note that these values are plotted logarithmically.}
\end{figure}

The final $f_{\star}$ map is created by taking the ratio of the
resolution-corrected stellar mass surface density map to the total mass 
surface density map created by our model; this map is shown in Figure 
\ref{fig:fstar}. To obtain the errors in $f_{\star}$, we resample the total 
mass surface density maps from our model, and create stellar mass surface 
density error maps by resampling from distributions of $M_{\star}/L$ ratios 
and ICL mass fractions. We use these to determine all errors in quantities 
derived from these maps.

We find $\langle f_{\star} \rangle = 0.012^{+0.004}_{-0.003}$ when we average
over a circular aperture, centered on the BCG, with a radius of 0.3 Mpc. 
If we use a Chabrier or Salpeter IMF, our $\langle f_{\star} \rangle$ 
values are $0.016^{+0.006}_{-0.005}$ (for a Salpeter IMF) or $0.010 \pm 0.003$ 
(for a Chabrier IMF). We
compare this value with other values for $\langle f_{\star} \rangle$ recently
obtained in the literature. Using the same SWUnited modeling package to
determine the total mass distribution, \citet{hoag16} produced maps for the
cluster MACS 0416.1-2403 ($z=0.397$, MACS0416 hereafter); using these maps 
and our updated stellar mass map smoothing procedure, we find 
$\langle f_{\star} \rangle = 0.005 \pm 0.001$ in a circle of radius 0.3 Mpc. 
Similarly, using the SWUnited maps of Abell 2744 ($z=0.308$, A2744 hereafter) 
created by \citet{wang16}, in a radius of 0.3 Mpc we find a value of 
$\langle f_{\star} \rangle = 0.003 \pm 0.001$. While these values are 
different than the original published values for $\langle f_{\star} \rangle$ 
due to our improved treatment of the stellar mass maps, they are still 
consistent with the original values within error bars. Notably, the model
of MACS1149 predicts twice as much total mass as
the models for MACS0416 or A2744, while all the clusters have comparable
stellar mass estimates.

In another study of MACS0416, \citet{jauz16} obtain a value of 
$\langle f_{\star} \rangle = 0.0315 \pm 0.0057$, using a smaller field of view 
and a Salpeter IMF rather than the diet-Salpeter IMF used in this work. If we
re-calculate our value of $\langle f_{\star} \rangle$ for MACS1149 using a
Salpeter IMF and the field of view used in \citet{jauz16}, we obtain an
average stellar to total mass ratio of $0.044^{+0.015}_{-0.012}$, consistent 
with the value obtained by \citet{jauz16} for a different, but similarly
complex, cluster. 

We also compare the distribution of stellar mass fraction across the
cluster. We find that $f_{\star}$ peaks toward the center of the cluster. This
suggests that the stellar mass surface density of MACS1149 peaks more sharply
near the cluster center than the total surface mass density, i.e., that the
stellar mass comprises a larger fraction of the total mass in the cluster
center than on its outskirts. This effect would be enhanced with use 
of a heavier IMF, radial gradients in IMFs due to heavier masses in the central,
dominant galaxies \citep{auger10}, and radial gradients in the ICL that 
concentrate in the center of the cluster \citep{morishita16b}. The value of 
$f_{\star}$ near the BCG is also very sensitive the procedure used for 
smoothing the stellar mass map \citep[see Figure 9,][]{hoag16}. Outside 
the center of the cluster, we find the mass-rich northwest association of 
galaxies, which demonstrates a strong dark matter component in the convergence 
map (Figure \ref{fig:kappa}), has a much lower $f_{\star}$ value than the other 
dark matter peaks, though the uncertainty of the total mass model in this region 
is much higher than in other regions due to relative lack of multiple images.

In addition to the inherent differences between clusters and possible
projection effects, we have better model constraints near the cluster center
for MACS1149 than were available for the analyses of A2744 and MACS0416
\citep{wang16, hoag16}. \citet{hoag16} demonstrated large systematic errors in
$f_{\star}$ associated with smoothing. The original \citet{wang16} analysis did
not undertake any smoothing of the stellar mass surface density map, and the
\citet{hoag16} analysis had few multiple images near the cluster center,
leading to large uncertainties in the smoothing in the center. In this work,
because of the presence of SN Refsdal and its highly-resolved spiral host
galaxy near the center of the cluster, we have a much higher resolution in
this region. Thus, our analysis improves the smoothing methods developed by
\citet{hoag16, wang16}, and we consider our conclusions about $f_{\star}$ to be
more robust, particularly in the center of the cluster.

In a study of baryon mass fractions in high-mass structures,
\citet{gonzalez13} measured $f_{\star}$ for twelve dynamically relaxed galaxy
clusters chosen from the X-ray sample of \citet{gonzalez07}, with redshifts
$z<0.3$. For clusters more massive than $2 \times 10^{14} M_{\odot}$, within an
aperture of $r_{500}$, they find a typical value of $\langle f_{\star} \rangle$
below 2\%, consistent with our results.

\citet{gonzalez13} also discuss the stellar mass fraction relative to the
cosmic baryon fraction value ($\frac{\Omega_b}{\Omega_m}$) and the ratio of
stellar mass to gas mass ($\frac{f_{\star}}{f_g}$). Since most cluster baryon
mass is comprised of intracluster gas, these values may be used to understand
a cluster's integrated star formation efficiency. \citet{gonzalez13} find that
for clusters with total mass $\sim 10^{15} M_{\odot}$, the value of
$\frac{f_{\star}}{f_g}\approx 0.1$ out to a radius of $r_{500}$.
Using the value of $r_{500}$ and gas mass estimate of MACS1149 found by
\citet{mantz10} and the mass-to-light ratio assumed in \citet{gonzalez13},
we model the stellar mass out to $r_{500}$ ($1.53 \pm 0.08 \rm{Mpc}$,
$4.00 \pm 0.21 \farcm$) by assuming a Gaussian distribution of stellar mass
surface density over large scales, centered on the peak stellar mass surface
density.  We find a stellar mass fraction of $\sim 2\%$, and
a value of $\frac{f_{\star}}{f_g}\approx 0.16$. These values are broadly
consistent with the results of \citet{gonzalez13}. We also calculate a
cluster baryon fraction value of $\frac{\Omega_b}{\Omega_m}\approx 13\%$,
and compare to the Planck value of $\frac{\Omega_b}{\Omega_m}=15.6\pm3.6\%$.
\citep{planck}

Our results are also consistent with those of \citet{ascaso14}, who used a
method that does not rely on lensing and a Chabrier IMF to establish
$f_{\star} = 0.00426$ at $z=0.1$ and $f_{\star} = 0.00398$ at $z=1$, for
$R<R_{\rm{vir}}$. This suggests that the integrated star formation efficiency
for MACS1149 is within the typical range for massive clusters, and within the
uncertainties is consistent with a relaxed cluster with a typical star
formation history integrated from $z_{\rm{formation}}$. 

\section{Conclusions}
The galaxy cluster MACS1149 is a complex merging cluster with a multiply
lensed spiral galaxy hosting a supernova. We use 38 independent multiple
images of background galaxies corresponding to 13 total systems, and an
additional 100 images corresponding to 31 star-forming regions in the 
SN Refsdal host galaxy. Using these constraints and an adaptive grid 
reconstruction method that combines strong and weak lensing, we
produce a mass model of this cluster, which yields critical curves with
a general shape that closely agrees with those of other current cluster lens
models of MACS1149.

We present a stellar mass surface density map using
\textit{Spitzer}/IRAC data, and compare this map to the total mass density
obtained from our lens model to obtain a map of the projected stellar mass to
total mass ratio, $f_{\star}$. We find that for MACS1149, stellar mass tends
to be more concentrated toward the cluster BCG, compared to the distribution
of total mass (stellar + gas + dark matter). We obtain a mean stellar mass
fraction $\langle f_{\star} \rangle = 0.012^{+0.004}_{-0.003}$ within 0.3 Mpc 
of the BCG, which is somewhat higher than the average stellar mass fraction of 
two other HFF clusters studied (MACS0416 and A2744), but broadly consistent 
with other estimates of this ratio. Discrepancies between the values of 
$f_{\star}$ between clusters can be explained by differences in smoothing 
techniques used, as well as by the natural variation between clusters. 
Moreover, this stellar mass fraction is consistent with the range of integrated 
star formation efficiency values for massive clusters of $z<1$. 

Additionally, both the mass model and the $f_{\star}$ map provide 
valuable clues for the ongoing discussions of how stellar light traces mass. 
We find that, in general, the stellar light in MACS1149 traces its mass, with a 
substantial dark matter peak near the cluster BCG. While a single cluster 
cannot on its own speak to these larger theoretical questions, the results 
observed for MACS1149 can be combined with those of other clusters to yield 
insight into the LTM assumption, and the various theories that rest upon it.




\acknowledgments

We acknowledge PIs Diego, Grillo, Jauzac, Kawamata, Sharon, and Zitrin, who
provided results from their models to use in this work. Additional models are 
obtained from the Mikulski Archive for Space Telescopes (MAST). EQF would like 
to thank Ben Cain for helpful conversations concerning the SWUnited code and 
for providing general lens modeling tips and Brent Follin for discussion of 
the cosmic baryon fraction. Support for this work was provided by NRAO via the 
NRAO Student Observing Support (SOS) Program. Support for this work was also 
provided by NASA through an award issued by JPL/Caltech and through 
\facility{HST}-AR-13235, \facility{HST}-AR-14280 from STScI. Observations were 
carried out using {\it Spitzer} Space Telescope, which is operated by the Jet 
Propulsion Laboratory, California Institute of Technology under a contract 
with NASA and using the NASA/ESA Hubble Space Telescope, obtained at the Space 
Telescope Science Institute, which is operated by the Association of Universities 
for Research in Astronomy, Inc., under NASA contract NAS 5-26555 and NNX08AD79G. 
AH acknowledges support by NASA Headquarters under the NASA Earth and Space 
Science Fellowship Program —-- Grant ASTRO14F-0007. The GLASS collaboration is 
supported by HST grant GO-13459.\\



Facilities: \facility{HST}, \facility{\emph{Spitzer}(IRAC)}, \facility{VLT(MUSE)}

\end{document}